\documentclass{webofc}
\usepackage[varg]{txfonts}   
\usepackage{graphics}
\usepackage{graphicx}
\usepackage{rotating}
\usepackage{epsfig}
\usepackage{bm}
\usepackage{longtable}
\begin{document}
\title{Heavy baryon spectroscopy}
%
%

\author{\firstname{Rudolf N.} \lastname{Faustov}\inst{1} \and
        \firstname{Vladimir O.} \lastname{Galkin}\inst{1}}

\institute{Institute of Cybernetics and  Informatics in Education, FRC
  CSC RAS, Moscow, Russia}

\abstract{%
  Masses of heavy baryons are calculated in the relativistic
  quark-diquark picture. Obtained results are in good agreement with available
  experimental data including recent measurements by the LHCb Collaboration. 
Possible quantum numbers of excited heavy baryon states are discussed.}
\maketitle
\section{Introduction}
\label{intro}
Recently significant experimental progress has been achieved in
studying heavy baryon spectroscopy. Many new heavy baryon states have
been observed. The main contribution was made by the LHCb
Collaboration. Thus last year the amplitude analysis of the decay
$\Lambda^0_b \to D^0p\pi^-$ was performed in the region of the phase
space containing $D^0p$ resonant contributions which revealed three
$\Lambda_c$ excited states and allowed to measure precisely their
masses and decay widths \cite{LHCb1}: the $\Lambda_c(2880)^+$ with the preferred
spin $J=5/2$; the new state $\Lambda_c(2860)^+$ with quantum numbers
$J^P=3/2^+$, its parity was measured relative to that of the
$\Lambda_c(2880)^+$; the $\Lambda_c(2940)^+$ with the  most likely spin-parity
assignment $J^P=3/2^-$ but other solutions with spins from $1/2$ to $7/2$
were not excluded. Then five new, narrow excited $\Omega_c$
states decaying to $\Xi_c^+ K^-$ were observed \cite{LHCb2}: the
$\Omega_c(3000)^0$, $\Omega_c(3050)^0$, $\Omega_c(3066)^0$,
$\Omega_c(3090)^0$, and  $\Omega_c(3119)^0$. These states were later
confirmed by Belle \cite{Belle}. Soon the discovery of the
long-awaited doubly charmed baryon $\Xi_{cc}^{++}$ was reported
\cite{LHCb3}. This year the new $\Xi_b(6227)^-$ resonance was observed
as a peak in both the $\Lambda_b^0K^-$ and $\Xi_b^0\pi^-$ invariant
mass spectra \cite{LHCb4}. Finally, the first observation of two
structures $\Sigma(6097)^\pm$ consistent with resonances in the final
states $\Lambda_b^0\pi^-$ and $\Lambda_b^0\pi^+$ was reported by the
LHCb \cite{LHCb5}.

In this talk we compare these new data with the predictions of the
relativistic quark-diquark model of heavy baryons \cite{hbar,hbarregge,dhb}. 
\section{Relativistic quark-diquark model of heavy baryons}

Our approach is based on the relativistic quark-diquark picture and the quasipotential equation. The interaction of two quarks in a diquark and the quark-diquark interaction  in a baryon are described by the
diquark wave function $\Psi_{d}$ of the bound quark-quark state
and by the baryon wave function $\Psi_{B}$ of the bound quark-diquark
state respectively.  These wave functions satisfy the relativistic
quasipotential equation of the Schr\"odinger type \cite{hbar}
\begin{equation}
\label{quas}
{\left(\frac{b^2(M)}{2\mu_{R}}-\frac{{\bf
p}^2}{2\mu_{R}}\right)\Psi_{d,B}({\bf p})} =\int\frac{d^3 q}{(2\pi)^3}
 V({\bf p,q};M)\Psi_{d,B}({\bf q}),
\end{equation}
where $\mu_{R}$ is the relativistic reduced mass, $b^2(M)$ is the center-of-mass
relative momentum squared on the mass shell, ${\bf p, q}$ are the
off-mass-shell relative momenta,
and $M$ is the bound state mass (diquark or baryon).

The kernel 
$V({\bf p,q};M)$ in Eq.~(\ref{quas}) is the quasipotential operator of
the quark-quark or quark-diquark interaction which is constructed with
the help of the
off-mass-shell scattering amplitude, projected onto the positive
energy states. We assume that the effective
interaction is the sum of the usual one-gluon exchange term and the mixture
of long-range vector and scalar linear confining potentials, where
the vector confining potential contains the Pauli term. The vertex of the 
diquark-gluon interaction takes into account the diquark
internal structure and effectively smears the Coulomb-like
interaction. The corresponding form factor is expressed as an overlap
integral of the diquark wave functions. Explicit
expressions for the quasipotentials of the quark-quark interaction in
a diquark and quark-diquark interaction in a baryon can be found in
Ref.~\cite{hbarregge}.
All parameters of the model 
were
fixed previously from considerations of meson  properties and are kept
fixed in the baryon spectrum calculations.

The quark-diquark picture of heavy  baryons reduces a very complicated
relativistic three-body problem to a significantly more simpler two
step two-body calculation. 
First we determine the properties of diquarks. We consider a diquark
to be a composite $(qq')$ system. Thus diquark in our approach is not a
point-like object. Its interaction with gluons is smeared by the form
factor expressed through the overlap integral of diquark wave
functions. These form factors enter the diquark-gluon interaction and
effectively take diquark structure into account \cite{hbarregge,dhb}. Note
that the ground state diquark composed from quarks with different
flavours can be both in scalar and axial vector state, while the
ground state diquarks composed from quarks of the same flavour can be
only in the axial vector state due to the Pauli principle.   Solving the
quasipotential equation numerically we  calculate the masses,
determine the diquark wave functions and use them for evaluation of
the diquark form factors. Only ground-state scalar
and axial vector diquarks are considered for heavy baryons.
While both qround-state as well as orbital and radial excitations of heavy
diquarks are necessary for doubly heavy baryons, since the lowest
excitations of such baryons originate from the
excitations of the  doubly heavy diquark.

Next we calculate the masses of heavy  baryons in the
quark--diquark picture \cite{hbarregge,dhb}. The heavy baryon
is considered as a bound state of a
heavy-quark and light-diquark. All excitations are assumed to occur
between heavy quark and light diquark. On the other hand,
the doubly heavy baryon is considered as a bound state of a
light-quark and heavy-diquark. Both excitations in the quark-diquark
system and excitations of the heavy diquark are taken into account.
It is important to note that such approach predicts significantly less
excited states of baryons compared to a genuine three-quark picture.
We do not expand the potential of the quark--diquark interaction
either in $p/m_{q,Q}$ or in $p/m_d$ and treat both diquark and 
quark fully relativistically.

\section{Heavy baryons}
\label{sec:hb}

The calculated masses of heavy  baryons are
given in Tables~\ref{tab:lq}-\ref{tab:oq}. In the first column we show the baryon
total isospin $I$, spin $J$ and parity 
$P$. The second column lists the quark-diquark state. The next
three columns refer to the charm and the last there columns to
the bottom baryons. There we first give our prediction for the
mass, then available experimental data \cite{pdg}: baryon status and
measured mass. The charm and bottom baryon states  recently discovered by
the LHCb Collaboration \cite{LHCb1,LHCb2,LHCb3,LHCb4,LHCb5} are marked as new.

\begin{table}
  \centering
\caption{ Masses  of the $\Lambda_Q$ ($Q=c,b$) heavy
  baryons (in MeV).}\label{tab:lq}
\begin{tabular}{ccccc|cccc}
  \hline
& & \multicolumn{3}{c|}{$Q=c$}& \multicolumn{3}{c}{$Q=b$}\\
\cline{3-5} \cline{6-8}
$I(J^P)$& $Qd$ state & $M$ &status& $M^{\rm exp}$   &  $M$ &status&
$M^{\rm exp}$ \\[2pt]
\hline
$0(\frac12^+)$& $1S$ & 2286&**** & 2286.46(14) & 5620&*** & 5619.51(23)\\
& $2S$ &{ 2769}&* &{ 2766.6(2.4)?}& 6089 \\
& $3S$ & 3130 &  &         & 6455  \\
& $4S$ & 3437 &  &         & 6756  \\
& $5S$ & 3715 &  &         & 7015  \\
& $6S$ & 3973 &  &         & 7256  \\
\hline
$0(\frac12^-)$& $1P$ & 2598 &***& 2592.25(28) & 5930 &***&5912.11(26) \\
& $2P$ &{ 2983}&*** & {2944.8$(^{1.4}_{1.5})$?} & 6326  \\
& $3P$ & 3303 &  &         & 6645  \\
& $4P$ & 3588 &  &         & 6917  \\
& $5P$ & 3852 &  &         & 7157  \\
\hline
$0(\frac32^-)$& $1P$ & 2627 &***& 2628.1(6) & 5942 &***&5919.81(23) \\
& $2P$ & 3005 &  &         & 6333  \\
& $3P$ & 3322 &  &         & 6651  \\
& $4P$ & 3606 &  &         & 6922  \\
& $5P$ & 3869 &  &         & 7171  \\
\hline
$0(\frac32^+)$& $1D$ &{ 2874} &new  & {2856.1.3$(^{2.3}_{5.9})$}        & 6190  \\
& $2D$ & 3189 &  &         & 6526  \\
& $3D$ & 3480 &  &         & 6811  \\
& $4D$ & 3747 &  &         & 7060  \\
\hline
$0(\frac52^+)$& $1D$ & 2880 &***& 2881.75(35)& 6196  \\
& $2D$ & 3209 &   &        & 6531 \\
& $3D$ & 3500 &  &         & 6814 \\
& $4D$ & 3767 &  &         & 7063 \\
\hline
$0(\frac52^-)$& $1F$ & 3097 &  &         & 6408  \\
& $2F$ & 3375 &  &         & 6705  \\
& $3F$ & 3646 &  &         & 6964  \\
& $4F$ & 3900 &  &         & 7196  \\
\hline
$0(\frac72^-)$& $1F$ & 3078 &  &         & 6411  \\
& $2F$ & 3393 &  &         & 6708  \\
& $3F$ & 3667&   &        & 6966  \\
& $4F$ & 3922 &  &         & 7197  \\
\hline
$0(\frac72^+)$& $1G$ & 3270 &  &         & 6598  \\
& $2G$ & 3546 &  &         & 6867  \\
\hline
$0(\frac92^+)$& $1G$ & 3284 &  &         & 6599  \\
& $2G$ & 3564 &  &         & 6868  \\
\hline
$0(\frac92^-)$& $1H$ & 3444 &  &         & 6767  \\
\hline
$0(\frac{11}2^-)$& $1H$ & 3460 & &          & 6766  \\
\hline
\end{tabular}
\end{table}

\begin{table}
\caption{Masses of the $\Sigma_Q$ ($Q=c,b$) 
  heavy baryons (in MeV).}\label{sigq}
\centering
  \begin{tabular}{ccccc|cccc}
\hline
& & \multicolumn{3}{c|}{$Q=c$}& \multicolumn{3}{c}{$Q=b$}\\
\cline{3-5} \cline{6-8}
$I(J^P)$& $Qd$ state & $M$ &status& $M^{\rm exp}$   &  $M$ &status&
$M^{\rm exp}$ \\
\hline
$1(\frac12^+)$& $1S$ & 2443 &****& 2453.76(18) & 5808 &***&
5807.8(2.7)\\
& $2S$ & 2901 & &           & 6213&  \\
& $3S$ & 3271 & &           & 6575& \\
& $4S$ & 3581 & &           & 6869&  \\
& $5S$ & 3861 & &           & 7124&  \\
\hline
$1(\frac32^+)$& $1S$ & 2519 &***& 2518.0(5)  & 5834 &***&
5829.0(3.4)\\
& $2S$ &{ 2936}&*** & { 2939.3$(^{1.4}_{1.5})$?} &
6226 &\\
& $3S$ & 3293 & &           &  6583 &\\
& $4S$ & 3598 & &           &  6876 &\\
& $5S$ & 3873 & &           &  7129 &\\
\hline
$1(\frac12^-)$& $1P$ & 2799 &***& 2802($^4_7$)& 6101 &     \\
& $2P$ & 3172 & &           &  6440 & \\
& $3P$ & 3488 & &           &  6756 & \\
& $4P$ & 3770 & &           &  7024 & \\
& $1P$ & 2713 & &           &  6095 &    \\
& $2P$ & 3125 & &           &  6430 & \\
& $3P$ & 3455 & &           &  6742 & \\
& $4P$ & 3743 & &           &  7008 & \\
\hline
$1(\frac32^-)$& $1P$ & 2798 &***& 2802($^4_7$)&  6096 & new& 6095.8(1.8)\\
& $2P$ & 3172 & &           &  6430 & &\\
& $3P$ & 3486 & &           &  6742 & \\
& $4P$ & 3768 & &           &  7009 & \\
& $1P$ &{2773}&* &{ 2766.6(2.4)?}& 6087 & \\
& $2P$ & 3151 & &           &  6423 & \\
& $3P$ & 3469 & &           &  6736 & \\
& $4P$ & 3753 & &           &  7003 & \\
\hline
$1(\frac52^-)$& $1P$ & 2789 & &           &  6084 & \\
& $2P$ & 3161 & &           &  6421 & \\
& $3P$ & 3475 & &           &  6732 & \\
& $4P$ & 3757 & &           &  6999 & \\
\hline
$1(\frac12^+)$& $1D$ & 3041 & &           &  6311 & \\
& $2D$ & 3370 & &           &  6636 & \\
\hline
$1(\frac32^+)$& $1D$ & 3043 & &           &  6326 & \\
& $2D$ & 3366 & &           &  6647 & \\
& $1D$ & 3040 & &           &  6285 & \\
& $2D$ & 3364 & &           &  6612 & \\
\hline
$1(\frac52^+)$& $1D$ & 3038 & &           &  6284 & \\
& $2D$ & 3365 & &           &  6612 & \\
& $1D$ & 3023 & &           &  6270 & \\
& $2D$ & 3349 & &           &  6598 & \\
\hline
$1(\frac72^+)$& $1D$ &  3013& &            &  6260 & \\
& $2D$ & 3342 & &           &  6590 & \\
    \hline
    \end{tabular}
\end{table}

\begin{table}[hbt]
  \centering
\caption{ Masses of the $\Xi_Q$ ($Q=c,b$) heavy baryons
  with the scalar diquark (in MeV).}\label{tab:sqs}
  \begin{tabular}{ccccc|cccc}
\hline
& & \multicolumn{3}{c|}{$Q=c$}& \multicolumn{3}{c}{$Q=b$}\\
\cline{3-5} \cline{6-8}
$I(J^P)$& $Qd$ state & $M$ &status& $M^{\rm exp}$   &  $M$ &status&
$M^{\rm exp}$ \\[2pt]
\hline
$\frac12(\frac12^+)$& $1S$ & 2476 &***& 2470.88$(^{34}_{80})$   & 5803& ***&5790.5(2.7) \\
& $2S$ & 2959 & &            & 6266  \\
& $3S$ & 3323 & &            & 6601  \\
& $4S$ & 3632 & &            & 6913  \\
& $5S$ & 3909 & &            & 7165  \\
\hline
$\frac12(\frac12^-)$& $1P$ & 2792 &***& 2792.8(1.2) & 6120  \\
& $2P$ & 3179 & &            & 6496  \\
& $3P$ & 3500 & &            & 6805  \\
& $4P$ & 3785 & &            & 7068  \\
&$5P$ & 4048 & &            & 7302  \\
\hline
$\frac12(\frac32^-)$& $1P$ & 2819 &***  &2820.22(32) & 6130  \\
& $2P$ & 3201 & &            & 6502  \\
& $3P$ & 3519 & &            & 6810  \\
& $4P$ & 3804 & &            & 7073  \\
& $5P$ & 4066 & &            & 7306  \\
\hline
$\frac12(\frac32^+)$& $1D$ & 3059 &***&3055.9(0.4)  & 6366  \\
& $2D$ & 3388 & &            & 6690  \\
& $3D$ & 3678 & &            & 6966  \\
    & $4D$ & 3945 & &            & 7208  \\
    \hline
$\frac12(\frac52^+)$   & $1D$ & 3076 &* &3079.9(1.4) & 6373  \\
& $2D$ & 3407 &           &  & 6696 \\
& $3D$ & 3699 &           &  & 6970 \\
& $4D$ & 3965 &           &  & 7212 \\
\hline
 \end{tabular}
\end{table}

From Tables~\ref{tab:lq}, \ref{sigq}  we see that the
$\Lambda_c(2765)$ (or $\Sigma_c(2765)$), 
if it is indeed the $\Lambda_c$
state,  can be interpreted in our model as the first radial ($2S$)
excitation of the $\Lambda_c$. If instead it is the $\Sigma_c$ state, then it can
be identified as its first orbital excitation ($1P$) with
$J=\frac32^-$ (see Table~\ref{sigq}). The $\Lambda_c(2880)$ baryon
corresponds to the second orbital excitation ($2D$) with $J=\frac52^+$ in accord
with the LHCb analysis \cite{LHCb1}. The other charmed baryon, denoted as
$\Lambda_c(2940)$, probably has $I=0$, since it was discovered in the $pD^0$ mass
spectrum and not observed in $p D^+$ channel, but $I=1$ is not ruled out \cite{pdg}. If
it is really the $\Lambda_c$ state, then it could be both an orbitally and
radially excited ($2P$) state with $J=\frac12^-$, whose mass is
predicted to be about 40 MeV heavier. A better agreement with
experiment (within few MeV) is achieved, if the $\Lambda_c(2940)$ is
interpreted as the first radial excitation ($2S$) of the $\Sigma_c$
with $J=\frac32^+$. The $\Sigma_c(2800)$ can be identified with one of
the first orbital ($1P$) excitations of the $\Sigma_c$  with
$J=\frac12^-$ or $\frac32^- $ which have very close masses compatible
with experimental value within errors (see Table~\ref{sigq}). The
new state $\Lambda_c(2860)$ with quantum numbers $\frac32^+$ \cite{LHCb1}
can be well interpreted as second orbital excitation ($2D$ state). In
the bottom sector the $\Lambda_b(5912)$ and $\Lambda_b(5920)$
correspond to the first orbitally excited ($1P$) states with
$\frac12^-$ and $\frac32^-$, respectively. The new $\Sigma_b(6097)$
state \cite{LHCb5} can be the first orbital excitation ($1P$) with
quantum numbers $\frac32^-$.

\begin{table}[hbt]
  \centering
\caption{ Masses of the $\Xi_Q$ ($Q=c,b$) heavy baryons
  with the axial vector diquark (in MeV).}\label{tab:sq}
  \begin{tabular}{ccccc|cccc}
\hline
& & \multicolumn{3}{c|}{$Q=c$}& \multicolumn{3}{c}{$Q=b$}\\
\cline{3-5} \cline{6-8}
$I(J^P)$& $Qd$ state & $M$ &status& $M^{\rm exp}$   &  $M$ &status&
$M^{\rm exp}$ \\
\hline
    $\frac12(\frac12^+)$& $1S$ & 2579&*** & 2577.9(2.9) & 5936&***&5935.02(5) \\
& $2S$ & 2983 & &2971.4(3.3) & 6329 \\
& $3S$ & 3377 & &&             6687 \\
& $4S$ & 3695 & &&             6978 \\
& $5S$ & 3978 & & &            7229 \\
\hline
$\frac12(\frac32^+)$& $1S$ & 2649 &***& 2645.9(0.5) & 5963&***&5955.33(13) \\
& $2S$ & 3026 &   &          & 6342 \\
& $3S$ & 3396 &   &          & 6695 \\
& $4S$ & 3709 &   &          & 6984 \\
& $5S$ & 3989 &    &         & 7234 \\
\hline
$\frac12(\frac12^-)$& $1P$ & 2936 & *&  2931(6)       & 6233&     \\
& $2P$ & 3313 &   &          & 6611 \\
& $3P$ & 3630 &   &          & 6915 \\
& $4P$ & 3912 &    &         & 7174 \\
& $1P$ & 2854 &    &         & 6227  & new &6226.9(2.1)   \\
& $2P$ & 3267 &    &        & 6604 \\
& $3P$ & 3598 &    &        & 6906 \\
& $4P$ & 3887 &    &         & 7164 \\
\hline
$\frac12(\frac32^-)$& $1P$ & 2935 & *& 2931(6)   & 6234 &  \\
& $2P$ & 3311 &  &           & 6605 \\
& $3P$ & 3628 &  &           & 6905 \\
& $4P$ & 3911 &  &          & 7163 \\
& $1P$ & 2912 &  &           & 6224 & new &6226.9(2.1) \\
& $2P$ & 3293 &  &           & 6598  \\
& $3P$ & 3613 &  &           & 6900  \\
& $4P$ & 3898 &  &          & 7159  \\
\hline
$\frac12(\frac52^-)$& $1P$ & 2929 & *&2931(6)    & 6226& new &6226.9(2.1)  \\
& $2P$ & 3303 & &            & 6596\\
& $3P$ & 3619 & &           & 6897\\
& $4P$ & 3902 & &           & 7156\\
\hline
$\frac12(\frac12^+)$& $1D$ & 3163 & &            & 6447  \\
\hline
$\frac12(\frac32^+)$& $1D$ & 3167 & &            & 6459 \\
& $1D$ & 3160 & &           & 6431 \\
\hline
$\frac12(\frac52^+)$& $1D$ & 3166 & &            & 6432 \\
& $1D$ & 3153 &  &          & 6420 \\
\hline
$\frac12(\frac72^+)$& $1D$ & 3147&* & 3122.9(1.3)   & 6414 \\
                 
\hline
 \end{tabular}
\end{table}

\begin{table}[hbt]
  \centering
\caption{ Masses of the $\Omega_Q$ ($Q=c,b$) heavy
  baryons (in MeV).}\label{tab:oq}
  \begin{tabular}{ccccc|cccc}
\hline
& & \multicolumn{3}{c|}{$Q=c$}& \multicolumn{3}{c}{$Q=b$}\\
\cline{3-5} \cline{6-8}
$I(J^P)$& $Qd$ state & $M$ &status& $M^{\rm exp}$   &  $M$ &status&
$M^{\rm exp}$ \\[2pt]
\hline
$0(\frac12^+)$& $1S$ & 2698 &***& 2695.2(1.7) & 6064&***& 6046.4(1.9)\\
& $2S$ & 3088 &new& {3090.2$(^7_8)$} & 6450 \\
& $3S$ & 3489 &  &           & 6804 \\
& $4S$ & 3814 &  &           & 7091 \\
& $5S$ & 4102 &  &           & 7338 \\
\hline
$0(\frac32^+)$& $1S$ & 2768 &***& 2765.9(2.0) & 6088\\
& $2S$ & 3123 &new&{3119.1$(^{1.0}_{1.1})$} & 6461 \\
& $3S$ & 3510 &  &           & 6811 \\
& $4S$ & 3830 &  &           & 7096 \\
& $5S$ & 4114 &  &           & 7343 \\
\hline
$0(\frac12^-)$& $1P$ & 3055 && & 6339     \\
& $2P$ & 3435 &  &           & 6710 \\
& $3P$ & 3754 &  &           & 7009 \\
& $4P$ & 4037 &  &           & 7265 \\
& $1P$ & 2966 && & 6330   \\
& $2P$ & 3384 &  &           & 6706 \\
& $3P$ & 3717 &  &           & 7003 \\
& $4P$ & 4009 &  &           & 7257 \\
\hline
$0(\frac32^-)$& $1P$ & 3054 &new&{ 3065.6$(^6_7)$}   & 6340 \\
& $2P$ & 3433 & &            & 6705 \\
& $3P$ & 3752 & &            & 7002 \\
& $4P$ & 4036 & &            & 7258 \\
& $1P$ & 3029 & new&{3000.4$(^4_6)$ }           & 6331\\
& $2P$ & 3415 & &            & 6699 \\
& $3P$ & 3737 & &            & 6998 \\
& $4P$ & 4023 & &            & 7250 \\
\hline
$0(\frac52^-)$& $1P$ & 3051 &new&{3050.2$(^4_5)$}  & 6334  \\
& $2P$ & 3427 & &            & 6700 \\
& $3P$ & 3744 & &            & 6996 \\
& $4P$ & 4028 & &            & 7251 \\
\hline
$0(\frac12^+)$& $1D$ & 3287 &  &           & 6540 \\
\hline
$0(\frac32^+)$& $1D$ & 3298 &  &           & 6549  \\
& $1D$ & 3282 &  &           & 6530 \\
\hline
$0(\frac52^+)$& $1D$ & 3297 &    &         & 6529 \\
& $1D$ & 3286 &    &         & 6520  \\
\hline
$0(\frac72^+)$& $1D$ & 3283 &    &         & 6517  \\
\hline
$0(\frac32^-)$& $1F$ & 3533 &   &           & 6763  \\
\hline
\end{tabular}
\end{table}

In the $\Xi_Q$ baryon sector as we see from Tables~\ref{tab:sqs},\ref{tab:sq} the
$\Xi_c(2790)$ and $\Xi_c(2815)$ can be assigned to the first orbital
($1P$) excitations of the $\Xi_c$ containing a scalar diquark with $J=\frac12^-$
and  $J=\frac32^-$, respectively. On the other hand, the charmed
baryon $\Xi_c(2930)$ can be considered as either the $J=\frac12^-$,
$J=\frac32^-$  or $J=\frac52^-$ state (all these states are predicted to have close masses)
corresponding to  the first orbital ($1P$) excitations of the $\Xi_c'$
with an axial vector diquark.  While the $\Xi_c(2980)$ can be viewed as  the
first radial ($2S$) excitation with $J=\frac12^+$ of the $\Xi_c'$, the
$\Xi_c(3055)$ and  $\Xi_c(3080)$ baryons can be interpreted  as a second
orbital ($2D$) excitations of the $\Xi_c$ containing a scalar diquark
with $J=\frac32^+$ and  $J=\frac52^+$, and the $\Xi_c(3123)$ can be
viewed as the corresponding ($2D$) excitation of the $\Xi_c'$  with
$J=\frac72^+$.  The recently observed excited bottom baryon
$\Xi_b(6227)^-$  \cite{LHCb4} can be one of the first radially excited
states ($1P$) of the  $\Xi_b'$ baryon with the axial vector diquark and quantum numbers
$\frac12^-$, $\frac32^-$, $\frac52^-$ which are predicted to have very
close masses.

\begin{figure}
\centerline{\includegraphics[width=8cm]{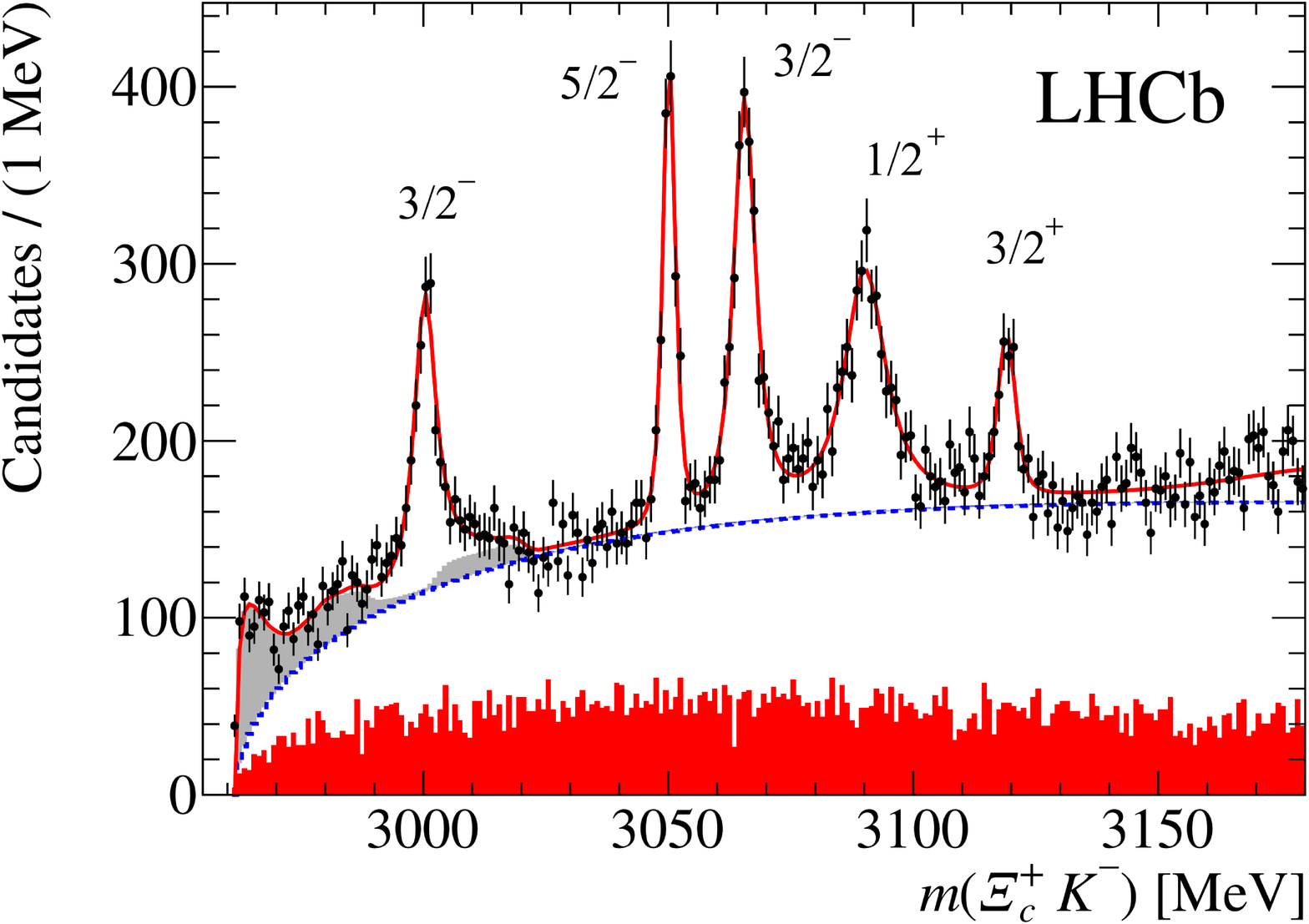}}
\caption{Proposed assignment of spins and parities of excited
  $\Omega_c$ states observed by LHCb Collaboration.}\label{fomc} \vspace*{-0.8cm} 
\end{figure}

Masses of the $\Omega_c$ and $\Omega_b$ baryons are given in
Table~\ref{tab:oq}. The ground state ($1S$) masses were predicted \cite{hbar}
before experimental discovery and agree well with measured values.
Recently observed \cite{LHCb2} five new, narrow excited $\Omega_c$
are also in accord with our predictions. Three lighter states
$\Omega_c(3000)^0$, $\Omega_c(3050)^0$ and $\Omega_c(3066)^0$ are well
described as first orbital ($1P$) excitations with $J=$ $\frac32^-$,
$\frac52^-$ and $\frac32^-$, respectively. These states are expected
to be narrow. The remaining $1P$ states with $\frac12^-$ are expected
to be broad and thus can escape detection. The small peak in the low
end of $\Xi_c^+K^-$ mass distribution (see Fig.~\ref{fomc}) can
correspond to $\frac12^-$ state with the predicted mass 2966~MeV (see
Table~\ref{tab:oq}). The remaining two heavier states 
$\Omega_c(3090)^0$ and  $\Omega_c(3119)^0$ are naturally described as
first radial ($2S$) excitations with quantum numbers  $\frac12^+$ and
$\frac32^+$, respectively. Their predicted masses coincide with the
measured ones within a few MeV. The proposed assignment of spins and
parities of excited  $\Omega_c$ states observed by LHCb Collaboration
is given in Fig.~\ref{fomc}. In Table~\ref {omcomp} we compare
different quark model (QM), QCD sum rules (QCD SR), lattice QCD predictions  and
available experimental data for the masses of the $\Omega_c$ states.

\begin{table}[hbt]
\caption{Comparison of theoretical predictions for the masses of the
  $\Omega_c$ states.}\label{omcomp}
\centering
\begin{tabular}{ccccccc c}
\hline
State & our \cite{hbarregge}&\cite{Rob} & \cite{Shah}&\cite{Bali} &\cite{Chen} & \cite{Ag} & Experiment. \\
$nL,J^P$&RQM &QM& QM & lattice &  lattice & QCD SR &  PDG+LHCb \\ \hline
$1S,\frac12^+$ & 2698&2718 & 2695&2648(28) & $2695(28)$ & $ 2685(123)$ & $2695.2(1.7)$ \\
$2S,\frac12^+$ & 3088&3152 & 3100 &3294(73)& & $3066(138)$  &  {3090.2$(^7_8)$} \\
$1S,\frac32^+$ & 2768&2776 & 2767 &2709(32)& $2781(25)$  & $2769(89)$ & $2765.9(2.0)$ \\
$2S,\frac32^+$ & 3123&3190 & 3126 &3355(92)&    & $3119(114)$  &{3119.1$(^{1.0}_{1.1})$}\\
$1P,\frac12^-$ & 2966&2977 & 3028 &2995(46)& $3015(45)$ & &  \\
$1P,\frac12^-$ & 3055&2990 & 3011 & & &  \\
$1P,\frac32^-$ & 3054&2986 & 2976 &3016(69)&  &  & {3065.6$(^6_7)$} \\
$1P,\frac32^-$ & 3029&2994 & 2993 &  &  &&{3000.4$(^4_6)$}\\
$1P,\frac52^-$ & 3051&3014 & 2947 &&  &  & { 3050.2$(^4_5)$} \\
\hline
\end{tabular}
\end{table}

\section{Doubly heavy baryons}

\label{sec:dhb}

\begin{table}
   \centering
\caption{\label{ccq}Mass spectrum of $\Xi_{cc}$ baryons (in GeV).}
\begin{tabular}{cccccc}
  \hline
State &  \multicolumn{2}{c}{Mass}  & State & \multicolumn{2}{c}{Mass} \\
$(n_dLn_ql)J^P$& our & \cite{Ger}&$(n_dLn_ql)J^P$& our &
\cite{Ger}\\ 
\hline
$(1S1s)\frac12^+$& 3.620& 3.478&$(1P1s)\frac12^-$  &3.838   &3.702   \\
$(1S1s)\frac32^+$& 3.727& 3.61 &$(1P1s)\frac32^-$  &3.959   &3.834    \\
$(1S1p)\frac12^-$& 4.053& 3.927&$(2S1s)\frac12^+$  &3.910   & 3.812  \\
$(1S1p)\frac32^-$& 4.101& 4.039&$(2S1s)\frac32^+$  &4.027   &3.944   \\
$(1S1p)\frac12'^-$& 4.136& 4.052&$(2P1s)\frac12^-$  &4.085   &3.972   \\
$(1S1p)\frac52^-$& 4.155& 4.047& $(2P1s)\frac32^-$ &4.197   &4.104   \\
$(1S1p)\frac32'^-$& 4.196& 4.034&$(3S1s)\frac12^+$  &4.154   & 4.072  \\
\hline
\end{tabular}
\end{table}

\begin{table}
  \centering
\caption{\label{tab:gmass}Mass spectrum of ground states of doubly
  heavy baryons (in GeV).  $\{QQ\}$ denotes the diquark in the axial vector state
  and $[QQ]$ denotes diquark in the scalar state.}\label{dhbm}
\begin{tabular}{@{}c@{ }ccccccccc}
\hline
Baryon&
Quark&$J^P$&our&\cite{Ger}&\cite{Ronc}&\cite{Nar} &\cite{Mart} &\cite{Rob}& \cite{Kar}\\
      &content&    &\cite{dhb}&     && & &&\\   
\hline
$\Xi_{cc}$  &$\{cc\}q$&$\frac12^+$&3.620&3.478&3.66&3.69&3.510&3.676&3.627(12)\\
$\Xi_{cc}^*$&$\{cc\}q$&$\frac32^+$&3.727&3.61&3.74&&3.548&3.753&3.690(12)\\
$\Omega_{cc}$&$\{cc\}s$&$\frac12^+$&3.778&3.59&3.74&3.86&3.719&3.815\\
$\Omega_{cc}^*$&$\{cc\}s$&$\frac32^+$&3.872&3.69&3.826&&3.746&3.876\\
$\Xi_{bb}$  &$\{bb\}q$&$\frac12^+$&10.202&10.093&10.34 &10.16&10.130&10.340&10.162(12)\\
$\Xi_{bb}^*$ &$\{bb\}q$&$\frac32^+$&10.237&10.133&10.37&&10.144&10.367&10.184(12)\\
$\Omega_{bb}$&$\{bb\}s$&$\frac12^+$&10.359&10.18&10.37 &10.34&10.422&10.454\\
$\Omega_{bb}^*$&$\{bb\}s$&$\frac32^+$&10.389&10.20&10.40& &10.432&10.486\\
$\Xi_{cb}$  &$\{cb\}q$&$\frac12^+$&6.933&6.82&7.04& 6.96&6.792&7.011&6.914(13)\\
$\Xi'_{cb}$  &$[cb]q$&$\frac12^+$&6.963&6.85&6.99&&6.825&7.047&6.933(12) \\
$\Xi_{cb}^*$ &$\{cb\}q$&$\frac32^+$&6.980&6.90&7.06& &6.827&7.074&6.969(14)\\
$\Omega_{cb}$  &$\{cb\}s$&$\frac12^+$&7.088&6.91&7.09 &7.13&6.999&7.136\\
$\Omega'_{cb}$  &$[cb]s$&$\frac12^+$&7.116&6.93&7.06 &&7.022&7.165\\
$\Omega_{cb}^*$ &$\{cb\}s$&$\frac32^+$&7.130&6.99&7.12 &&7.024&7.187\\
\hline
\end{tabular}
\end{table}

Mass spectra of doubly heavy baryons was calculated in the
light-quark--heavy-diquark picture in Ref.~\cite{dhb}. The light quark
was treated completely relativistically, while the expansion in the
inverse heavy quark mass was used. Table~\ref{ccq} shows the $\Xi_{cc}$
 mass spectrum. Excitaions inside doubly heavy diquark and
light-quark--heavy-diquark bound systems are taken into account. We use the
notations $(n_dLn_ql)J^P$, where we first show the radial quantum
number of the diquark ($n_d=1,2,3\dots$) and its orbital momentum by a
capital letter ($L=S,P,D\dots$) , then the radial quantum number of the
light quark ($n_q=1,2,3\dots$) and its orbital momentum by a
lowercase letter ($l=s,p,d\dots$), and at the end the total angular
momentum $J$ and parity $P$ of the baryon. In
Table~\ref{dhbm} we compare different theoretical predictions for the
ground state masses of the doubly heavy baryons. Our prediction (2002)
for
the mass of the $\Xi_{cc}$ baryon \cite{dhb} excellently agrees  with its
mass recently measured (2017) by the LHCb Collaboration \cite{LHCb5}:
$$M^{\rm exp}(\Xi_{cc}^{++})=3621.40\pm0.72\pm0.27\pm0.14~{\rm MeV}.$$

\section{Conclusions}
\label{sec:conc}

Recent observations of excited charm and bottom baryons confirm predictions of
the relativistic heavy-quark--light-diquark model of heavy
baryons \cite{hbar,hbarregge}. The new state $\Lambda_c(2860)$ is in accord with the
predicted $1D$- state with $J^P=\frac32^+$. The experimentally
preferred  quantum numbers $J^P=\frac52^+$ of $\Lambda_c(2880)$ agree
with our assignment of this state to $1D$- state with
$J^P=\frac52^+$. The $\Lambda_b(5912)$ and $\Lambda_b(5920)$ are well described as the first orbitally excited ($1P$) states with
$\frac12^-$ and $\frac32^-$, respectively. The new $\Sigma_b(6097)$
state can be the first orbital excitation ($1P$) with
quantum numbers $\frac32^-$. The recently observed excited bottom baryon
$\Xi_b(6227)^-$  can be one of the first radially excited
states ($1P$) of the  $\Xi_b'$ baryon with the axial vector diquark
and quantum numbers $\frac12^-$, $\frac32^-$, $\frac52^-$ which are
predicted to have very close masses. Observation of five new narrow
$\Omega_c$ states in the mass range 3000-3200 MeV agrees with our
prediction of orbitally excited $1P$-states and radially excited
$2S$-states in this mass region: $\Omega_c(3000)$, $\Omega_c(3066)$,
$\Omega_c(3050)$ can be $1P$-states with $J^P=\frac32^-,\frac32^-,\frac52^-$
while $\Omega_c(3090)$ and $\Omega_c(3119)$ states are most likely the first
radially excited $2S$ states with $J^P=\frac12^+,\frac32^+$.

In the doubly heavy baryon sector, the mass of recently observed
$\Xi_{cc}^{++}$ baryon is in excellent  agreement with our prediction
made more than 15 years ago \cite{dhb}. Masses of ground state doubly charm
baryons are predicted to be in 3.5 - 3.9 GeV range. Masses of ground
state doubly bottom baryons are predicted to be in 10.1 - 10.5 GeV range. 
Masses of ground state bottom-charm baryons are predicted to be
in  6.8 - 7.2 GeV range.
Rich spectra of narrow excited states below strong decay thresholds
are expected.  We strongly encourage experimenters to search for
new excited states of heavy baryons and especially for doubly heavy
baryons.

We are grateful to I. Belyaev, D. Ebert, M. Ivanov, M. Karliner and A.
Martin for valuable
discussions and support. We thank the organizers of XXIV International Baldin Seminar on High Energy Physics Problems.

\end{document}